\documentstyle[12pt]{article}
\newcommand{\beq}{\begin{equation}}
\newcommand{\eeq}{\end{equation}}
\newcommand{\id}
 {i\kern.06em\hbox{\raise.25ex\hbox{$/$}\kern-.60em$\partial$}}

\newcommand{\bs}{/\kern-.52em b}
\newcommand{\qs}{/\kern-.52em s}
\newcommand{\D}{{\cal{D}}}
\newcommand{\dv}{\!d^3\!x\,}

\newcommand{\dd}
{\kern.06em\hbox{\raise.25ex\hbox{$/$}\kern-.60em$\partial$}}

\newcommand{\tr}{\mathop{\rm tr}\nolimits}
\begin{document}
\title{ Duality between Topologically Massive and Self-Dual models }

\author{
J.C. Le Guillou$^a$\thanks{Also at {\it Universit\'e de Savoie} and at
{\it Institut Universitaire de France}}\,,
E. F. Moreno$^b$\thanks{Supported by CUNY Collaborative Incentive Grant 991999},
C. N\'u\~nez$^c$ \\
and\\
F.A. Schaposnik$^c$\thanks{Investigador CICBA, Argentina}
\\
~
\\
{\normalsize\it
$^a$Laboratoire de Physique Th\'eorique ENSLAPP}
\thanks{URA 1436 du CNRS associ\'ee
\`a l'Ecole Normale Sup\'erieure de Lyon et \`a
l'Universit\'e de Savoie}\\
{\normalsize\it
LAPP, B.P. 110, F-74941 Annecy-le-Vieux Cedex, France}\\
~\\
{\normalsize\it
$^b$ Physics Department, City College of the City University
of New York}\\
{\normalsize\it
New York NY 10031, USA}\\
{\normalsize\it
Physics Department, Baruch College, The City University of
New York}\\
{\normalsize\it
New York NY 10010, USA}
 \\
~\\
{\normalsize\it
$^c$Departamento de F\'\i sica, Universidad Nacional de La Plata}\\
{\normalsize\it
C.C. 67, 1900 La Plata, Argentina}}
\date{}
\maketitle

\vspace{-5.5 in}

\hfill\vbox{
\hbox{\it ENSLAPP-A-655/97}
\hbox{\it CCNY-HEP 97/6}
\hbox{\it La Plata Th-16/97}
}
\vspace{4.5 in}

\begin{abstract}
{We show that, with the help of a general BRST symmetry, different
theories in 3 dimensions can be connected through a fundamental
topological field theory related to the classical limit of the
Chern-Simons model.}
\end{abstract}
\newpage
%



Different descriptions of massive self-dual vector fields in
three dimensions have been shown to exists both in the
Abelian and non-Abelian cases \cite{VN}-\cite{KLRV}. The action
is either linear in derivatives (the self-dual model) or
quadratic with a topological mass term \cite{DJT}.
The connection thus established
between different models was
recently exploited to analyze
relevant supersymmetric models \cite{KLRV} and also
to derive bosonization rules for three dimensional fermionic
systems \cite{FS}. A phase space analysis of
the connection in the abelian case can be found in
\cite{BRR}.

Trying to overcome some complications
that arise
in establishing the connection  in the non-Abe\-lian case
\cite{KLRV}  a ``large'' BRST invariance
was exploited in ref.\cite{LMNS} in connection
with non-Abelian bosonization in $d=3$ dimensions. It is the
purpose of this work to use this BRST invariance to show that
the different models
studied in \cite{VN}-\cite{KLRV} can all be described by
a unique partition function related to the
classical limit of the Chern-Simons theory partition
function.

~

Let us start by considering the following (Euclidean) path-integral
for a gauge field $b_\mu$ taking values in the Lie algebra
of some group $G$
\beq
Z = \sum_{\alpha}
\int \D b_\mu \delta[b_\mu -
{\tilde b}_\mu^{(\alpha)}] \,
\exp(\frac{\kappa}{2\pi} S_{CS}[b] )
\label{1}
\eeq
Here  $S_{CS}[b]$ is the Chern-Simons action,
\beq
  S_{CS}[b] =  i\varepsilon_{\mu\nu\lambda} \tr \int_{M}\dv
 (F_{\mu \nu}[b] b_{\lambda} -
   \frac{2}{3} b_{\mu}b_{\nu}b_{\lambda})  ,
\label{2}
\eeq
${\tilde b}_\mu^{(\alpha)}$ is a complete set
of gauge equivalence classes of flat connections and $M$ an
oriented three manifold.
Note that the constraint on the path-integration domain imposed
by the delta functional in eq.(\ref{1}) corresponds to saturate
the path-integral with the solution of the {classical} equations of
motion for the Chern-Simons action,
\beq
\varepsilon_{\mu\nu\lambda}F_{\mu \nu}[b] = 0 .
\label{4}
\eeq
In this sense, we could think of $Z$ as the partition function
for a Chern-Simons theory restricted to the classical sector.
With this restriction, the CS theory which is a Schwartz type topological
theory becomes a Witten type topological theory \cite{BBRT}. Indeed,
if one considers the most general transformation for $b_\mu$,
\beq
b_\mu \to b_\mu + \epsilon_\mu
\label{ex}
\eeq
which leaves invariant (Witten-type) topological Yang-Mills theory,
the CS action changes as
\beq
\delta_{\epsilon_{\mu}}S_{CS} = 2 i\varepsilon_{\mu \nu \alpha}
\int d^3x \epsilon_\mu F_{\nu \alpha}[b]
\label{cu}
\eeq
But, when  the $\delta$-function constraint is included as in (\ref{1}),
this non-invariant term vanishes. Then, the model defined
through partition function (\ref{1}) has a ``large'' topological
invariance as it occurs in
Witten type theories and, associated with it, it
 posses, as we shall see, a``large'' BRST invariance.
Now, let us note that integrating the delta function in
(\ref{1}) one just gets
\beq
Z = \sum_{\alpha}\exp(\frac{\kappa}{2\pi} S_{CS}[{\tilde b}^{(\alpha)}])
\label{co}
\eeq
We then see that the
partition function (\ref{1}) picks contributions solely from the
classical sector of the CS theory since only the phase of
 the CS invariant of the flat connection
$b^{(\alpha)}$, $S_{CS}[b^{(\alpha)}]$, appears in (\ref{co}).
This should be contrasted with  the ``complete'' Chern-Simons theory,
i.e. that with a partition function of the form
\beq
Z_{CS} =
\int \D b_\mu \exp(\frac{\kappa}{2\pi} S_{CS}[b] )
\label{1c}
\eeq
 Now, due to eq.(\ref{co}), it
is evident  that one can connect the theory defined by (\ref{1})
with the weak coupling limit of the complete CS theory since

\beq
\lim_{\kappa \to \infty}
\int \D b_\mu \exp(\frac{\kappa}{2\pi} S_{CS}[b] ) =
\lim_{\kappa \to \infty}
\sum_{\alpha}\exp(\frac{\kappa}{2\pi} S_{CS}[{\tilde b}^{(\alpha)}])
\label{fio}
\eeq
or, more compactly
\beq
\lim_{\kappa \to \infty} Z_{CS} =
\lim_{\kappa \to \infty}  Z
\label{fio1}
\eeq
That is, the limit of $\kappa \to \infty$ of our model coincides
with the same limit of the (complete) CS model.

According to the manifold,
$Z_{CS}$ can be evaluated using different
approaches \cite{Wi}-\cite{ILR}.  In particular,
for a manifold $M = S^2 \times S^1$, one has
$Z_{CS}[S^2 \times S^1,G] = 1 $
for any $G$ and any $k$, a result which can be
considered as a normalization definition for the
partition function of CS
theories in manifolds of the form $X \times S^1$.
For  $ M = S^3$ and $G = SU(2)$
 $Z_{CS}$ can also be evaluated, leading to
\beq
Z_{CS}[S^3,SU(2)] = \sqrt{\frac{2}{k+2}} \sin(\frac{\pi}{k+2})
\label{wi8}
\eeq
which behaves as $ Z \sim k^{-3/2} $ for large k. Finally,
 under some
circumstances (in particular, when there are no
ghost and auxiliary fields zero modes)
the partition function
for the  Chern-Simons theory can
be studied
in the weak coupling limit.
We thus see
that according to the manifold and the group, one
can compute the $k \to \infty$
limit of the CS partition function which can be identified
with our starting partition function (\ref{1}).


~

In order to make apparent the BRST invariance referred
above,  let us concentrate on an element of the equivalent class of 
flat gauge connections  (i.e. on a given term
in the sum defined in eq.(\ref{1})) and
write the delta functional of eq. (\ref{1}) in
the form \cite{LMNS}
\beq
\delta[b_\mu -  g^{-1} \partial_\mu g]  =
\vert \det (2 \varepsilon_{\mu \nu \alpha} D_\nu[b]) \vert
\delta [\varepsilon_{\mu\nu\lambda}F_{\mu \nu }[b]]
\label{6}
\eeq
where $D_\mu[b] = \partial_\mu + [b_\mu,~] $.
With this, the partition function (\ref{1}) can be rewritten as
\beq
Z  = \int \D b_\mu \D A_\mu \D \bar c_\mu \D c_\mu
 \exp (-S_{eff}[b,A, \bar c, c])
\label{10}
\eeq
where
\beq
S_{eff}[b,A, \bar c, c] =
-\frac{1}{2\pi} S_{CS}[b] - \frac{i}{\pi} \varepsilon_{\mu \nu \alpha}
tr\int \dv (A_\mu F_{\nu \alpha}[b] - 2\bar c_\mu D_\nu[b] c_\alpha) .
\label{11}
\eeq
Here $A_\mu$ is a one-form Lagrange multiplier introduced to
represent the delta functional and $\bar c_\mu$ and $c_\mu$ are
one-form ghost fields introduced to exponentiate the determinant in
(\ref{6}).(We have omitted a group volume factor in (\ref{10}).)
The partition function (\ref{10}) can be rewritten as
\beq
Z  = \int \D b_\mu \D A_\mu \D \bar c_\mu \D c_\mu \D d_\mu \D l \D \bar \xi
 \exp (-\tilde S_{eff}[b,A, \bar c, c, d, l, \bar \xi])
\label{13}
\eeq
where
\begin{eqnarray}
 & & \tilde S_{eff}[b,A, \bar c, c, d, l, \bar \xi]=
 - \frac{1}{2\pi} S_{CS}[b-d]
+\frac{i}{\pi} tr\!\!\int \dv
(-l d_\mu d_\mu + 2 \bar \xi c_\mu d_\mu)  \nonumber  \\
& & -
\frac{i}{\pi}\varepsilon_{\mu \nu \alpha} tr \!\!\int \dv
[(A_\mu + 2 d_\mu)F_{\nu \alpha}[b] - 2\bar c_\mu D_\nu[b] c_\alpha ]
\label{14}
\end{eqnarray}
Here $d_\mu$ is a one form auxiliary field taking values in the Lie
algebra, $l$ a scalar and $\bar \chi$ an antighost field. Integrating
out $l$ constrains $d_\mu$ to be zero so that (\ref{13}) is trivially
equivalent to (\ref{10}).
One can see that $\tilde S_{eff}$ possesses an (off-shell nilpotent)
BRST symmetry with transformations defined as
\[
  \delta b_\mu = c_\mu  \;\;\;\;\;\;\;\;\;\;\;\;\;\;\; \delta c_\mu = 0
\]
\beq
 \delta \bar c_\mu = A_\mu + 2 d_\mu \;\; \;\;\;\; \delta A_\mu = - 2
c_\mu \;\;\;\;\;\;  \delta d_\mu = c_\mu
\label{15}
\eeq
\[
\delta \bar \xi = l  \;\;\;\;\;\;\;\;\;\;\;\;\;\;\;\;\;\;\; \delta l = 0
\]
According to (\ref{15}),  $\tilde S_{eff}$ can be rewritten
in the form
\beq
 \tilde S_{eff} = - \frac{1}{2\pi} S_{CS}[b-d] -
\frac{i}{\pi} tr \int \dv
\delta   (\bar c_\mu {^*F}_\mu[b] + \bar \xi d_\mu d_\mu)
\label{16}
\eeq
and we have defined $^*F_\mu[b] = \varepsilon_{\mu \nu \alpha} F_{\nu
\alpha}[b]$. In view of the BRST invariance we can add  BRST exact
forms to $\tilde S_{eff}$, without changing the physics of the model.
If we  choose a functional $\;{\cal G}  = i (g^2/\pi) \int d^3x \; \bar
c_\mu A_\mu \;$ with $g$ an arbitrary constant carrying units of mass,
$[g^2 ] = m$ and add $\delta {\cal G}$ to the effective action one
generates a mass term for the $A_\mu$ field so that, after
integrating out $l$ giving $d_\mu = 0$, one finally ends with
\begin{eqnarray}
 \tilde S_{eff}[b,A,\bar c, c] & = &
 -\frac{1}{2\pi} S_{CS}[b] - \frac{i}{\pi} tr \int \dv A_\mu
 {^*F}_\mu[b]  +  \frac{i g^2}{\pi} tr  \int \dv A_\mu A_\mu +
 \nonumber \\ & & \frac{2i}{\pi} tr \int \dv \bar c_\mu (
 {^*D}_{\mu \alpha}[b] - {g^2} \delta_{\mu \alpha} ) c_\alpha
\label{20}
\end{eqnarray}
We thus see that $\tilde S_{eff}$ can be interpreted as the quantum
version of the ``master'' action introduced, in the Abelian case, in
refs.\cite{DJ} in order to establish the equivalence between the
massive self-dual theory and the topologically massive gauge theory
\cite{VN}. In the abelian case one can easily see this connection
starting from the master action and integrating out either $A_\mu$ or
$b_\mu$ \cite{FS}. Now, in the non-Abelian case, due to non-quadratic
terms appearing in the CS action, this exact integration is not
possible so that the equivalence between  the massive non-Abelian
self-dual model and Yang-Mills-Chern-Simons (or topologically
massive) theory cannot be proven in such a way.

More specifically,  starting from
\beq
Z = \int \D A \D b \D \bar c \D c \exp (-\tilde S_{eff}[b,A,\bar c,
c] )
\label{Z}
\eeq
one can easily perform the $A_\mu$ integration so that $Z$ takes the form
\begin{eqnarray}
Z & = &\int \D b \D \bar c \D c \exp( \frac{1}{2\pi}S_{CS}[b]
+\frac{i}{2\pi g^2} tr \int \dv F_{\mu \nu}^2 [b]) \times \nonumber
\\ & & \exp (-\frac{2i}{\pi} tr \int \dv \bar
c_\mu ( {^*D}_{\mu \alpha}[b] - {g^2} \delta_{\mu \alpha} ) c_\alpha
) .
\label{entra}
\end{eqnarray}
We thus see that, at least in the strong-coupling regime ($g^2 \to
\infty$), where the ghost integration trivially decouples, $Z$
corresponds to the partition function of topologically massive gauge
theory: $\lim_{g^2 \to \infty} Z = Z_{top}$.

Now, as stated above, interchanging the order of integration (first
integrating over $b_\mu$) does not lead to  a simple theory for the
$A$ field (as one does in the Abelian case). However, if exploiting
BRST invariance we add to $\tilde S_{eff}$  a BRST exact form $\delta
{\cal H}$, thus defining :
\beq
\hat S_{eff} = \tilde S_{eff}  -
\frac{2i}{\pi} \epsilon_{\mu \nu \alpha}\int d^3x \; \delta\{
\bar
  c_\mu\left([b_\nu,A_\alpha + 2 d_\alpha] -
 (b_\nu - d_\nu)(A_\alpha + 2 d_\alpha)
\right)\}
\label{sui}
\eeq
then, after integration over auxiliary fields,
non-quadratic terms accommodate in the form first written in
\cite{vN} so that the $b_\mu$ integration can be trivially factored
out. Indeed, $\hat S_{eff}$  takes the form
\begin{eqnarray}
& & \hat S_{eff}  =  -\frac{1}{2\pi} S_{CS}[b] +
\frac{i g^2}{\pi}tr \int \dv A_\mu A_\mu
- \frac{i}{\pi} tr \int \dv A_\mu ^*F_\mu[b] \nonumber \\
& &
 - \frac{2i}{\pi} (
 \varepsilon_{\mu \nu \alpha}tr  \int \dv A_\mu A_\nu b_\alpha
-
\bar c_\mu
( ^*D_{\mu \alpha}[{b+A}] - {g^2} \delta_{\mu \alpha} )
 c_\alpha ) .
\label{suis}
\end{eqnarray}
After performing the shift $b_\mu \to b_\mu - A_\mu$, the
 integration over $b_\mu$ and the ghost fields is factored out
leading to  the partition function of the Self-Dual model,
\beq
Z =  \int \D A_\mu \exp(-S_{SD}[A])
\label{fig}
\eeq
with
\beq
S_{SD}[A] = \frac{1}{2\pi} S_{CS}[A]
 +\frac{i g^2}{\pi}tr \int \dv A_\mu A_\mu ,
\label{figu}
\eeq
result which was not obtained in ref.\cite{vN}.

We have then established a new series of equivalencies which is
summarized in Figure 1. The original constrained Chern-Simons model
becomes, after addition of a BRST exact form $\delta {\cal G}$ an
effective model which, after integration over the Lagrange multiplier
$A_\mu$, corresponds to a class of topologically massive model. In
fact, in the $g^2 \to \infty$ limit it coincides with the
Yang-Mills-Chern-Simons theory originally introduced to topologically
generate a mass for the gauge field \cite{DJT}. If one still adds an
appropriate BRST variation the "integration over the field
$b_{\mu}$"
can be easily performed resulting in the partition function for a
self-dual model.


~


It is important at this point to analyze the issue of symmetries in
the different models we have connected. We started with a constrained
CS model endowed with gauge symmetry. Indeed, the theory defined by
partition function (\ref{1}) shows an invariance under the gauge
transformation
\beq
b_\mu \to h^{-1} b_\mu h +   h^{-1} \partial_\mu h
\label{ch1}
\eeq
\beq
g \to h g .
\label{ch2}
\eeq
This symmetry is maintained after introduction of the Lagrange multiplier
field $A_\mu$ and ghosts provided their transformation laws are of the form
\beq
A_\mu \to h^{-1} A_\mu h,   \;\;\;\;\;\;\;\;\;
c_\mu \to  h^{-1} c_\mu h,
\;\;\;\;\;\;\;\;\;
\bar c_\mu \to  h^{-1} \bar c_\mu h
\label{ch3}
\eeq
Now, after addition of the BRST exact form to pass from $\tilde
S_{eff}$ to $\hat S_{eff}$ this invariance is spoiled. It is then no
surprise that we arrive to a self-dual model in which the presence of
the CS action makes evident that transformation (\ref{ch3}), ensuring
that $A_\mu$ transforms covariantly and not as a gauge field, does
not leave invariant the action. Remarkably, one can overcome this
problem by introducing a supplementary vector field $V_\mu$ precisely
as advocated in ref.\cite{vN}.

Indeed, let us consider the following action, modification of $\tilde
S_{eff}$ by addition of the field $V_\mu$ taking values in the Lie
algebra of $G$,
\beq
S_{eff} = \tilde S_{eff}[b,A,\bar c,c,d, l,\bar \xi] +
  S[V-d] + \delta{\cal K} .
\label{s}
\eeq
Here  $S[V]$ is the action giving dynamics to the $V_\mu$ field and
${\cal K}$ will be chosen so as to establish connections, analogous
to those discussed above, but now in the presence of the vector field
$V_\mu$. Consider  the partition function
\beq
Z = \int \D \Phi \exp(- S_{eff})
\label{agre}
\eeq
where $\Phi$ represent  the complete set of fields now including
$V_\mu$. We impose to $V_\mu$ the BRST transformation law
\beq
\delta V_\mu = c_\mu
\label{maniana}
\eeq
and supplement symmetry transformations (\ref{ch3}) with
\beq
V_\mu \to g^{-1} V_\mu g +  g^{-1} \partial_\mu g .
\label{ch6}
\eeq
Let us consider a family of functionals ${\cal K}$,
depending on a parameter $\lambda$, defined as
\beq
{\cal K}[V,b,A,\bar c;\lambda] =  - \frac{i}{\pi} \varepsilon_{\mu
\nu \alpha}
\int d^3x
\left( \bar c_\mu[b_\nu - V_\nu,  A_\alpha] +
\lambda \bar c_\mu
( A_\nu + 2d_\nu)(A_\alpha + 2d_\alpha) \right)
\label{hoy2}
\eeq
One can easily see that the resulting effective action is invariant
(after integration over auxiliary fields) under gauge transformations
(\ref{ch3}), (\ref{ch6}). Indeed,  $S_{eff}$ reads, after integration
on auxiliary fields, :
\begin{eqnarray}
& & S_{eff}  =  -\frac{1}{2\pi} S_{CS}[b] +
\frac{i g^2}{\pi}tr \int \dv A_\mu A_\mu
- \frac{i}{\pi} tr \int \dv A_\mu ^*F_\mu[b]
+ S[V] \nonumber \\
& &
 - \frac{2i}{\pi} (
 \varepsilon_{\mu \nu \alpha}tr  \int \dv A_\mu A_\nu
 (b_\alpha - V_\alpha + \frac{\lambda }{2} A_\alpha )
- \bar c_\mu
( ^*D_{\mu \alpha}[V] - {g^2} \delta_{\mu \alpha} )
 c_\alpha ) 
\label{lacan}
\end{eqnarray}
which corresponds for $\lambda = 0$ to the quantum version of the
action eq. (11) of ref.(\cite{vN}).

Now, with the choice  $\lambda = 0$, $S_{eff}$  becomes quadratic in
$A_\mu$ and one can perform the path-integral thus obtaining
\beq
Z = \int \D b_\mu \D V_\mu  \exp(- S[b,V])
\label{agre2}
\eeq
with
\begin{eqnarray}
S[b,V]  & = &  -  \frac{1}{2\pi} S_{CS}[b] + S[V] + \log \det
(g^2 \delta_{\mu \alpha} - {^*D}_{\mu \alpha}[{V}]) \nonumber \\
& &
- \frac{i}{4\pi} tr \int \dv {^*F}_\mu[b] \; [g^2 \delta_{\mu \alpha}
- 2 {\varepsilon_{\mu \nu \alpha}} (b_\nu - V_\nu)]^{-1} \;
{^*F}_\alpha[b] \nonumber\\
 & &
+\frac{1}{2} \log \det [g^2 \delta_{\mu \alpha} - 2 {\varepsilon_{\mu
\nu \alpha}} (b_\nu - V_\nu)]
\label{cua6}
\end{eqnarray}
This expression corresponds to the quantum version of the
action discussed in ref.(\cite{vN}) (see eq.(13) of that paper).

If one instead tries to first integrate out $b_\mu$, cubic terms
prevent explicit integration. However, one can perform a change of
variables that factors out trivially this integration leaving the
exact partition function for the $A_\mu$ and $V_\mu$ fields. Indeed,
if we define $ b'_\mu = b_\mu + A_\mu$ the effective action reads, in
terms of the new variable $b'$ as
\beq
S_{eff} = -\frac{1}{2\pi}S_{CS}[b'] + {\hat S}[A,V]
\label{sui2}
\eeq
where
\begin{eqnarray}
& & {\hat S}[A,V] =  \frac{i}{\pi} \varepsilon_{\mu \nu \alpha}
tr \int \dv
(A_\mu D_\nu[V] A_\alpha +
(\frac{2}{3} - \lambda) A_\mu A_\nu A_\alpha )
  \nonumber \\
& & + \frac{i g^2}{\pi} tr \int \dv A_\mu A_\mu + S[V] +
\log \det (g^2 \delta_{\mu \alpha} - {^*D}_{\mu \alpha}[{V}])
\label{piu}
\end{eqnarray}
At the partition function level, the $b'_\mu$ integral trivially
factors out so that ${\hat S}[A,V])$ determines the quantum theory
which, for $\lambda = 1/3$ , corresponds to the first order action
discussed in \cite{vN} (see eq.(9) in this reference). Finally, for
$\lambda = 2/3$ we obtain an action $S[A,V]$ which makes contact
with the starting action of eq. (1) in that paper, describing the
dynamics of a massive self-dual non-Abelian gauge field.

We have thus shown both the quantum version of the dualization of the
ref. \cite{vN}, dualization which was only achieve in ref. \cite{vN}
using in fact a modified version of its eq. (1), and the existence of
a quantum dualization using strictly this eq. (1) at the classical
level.

This extends the picture we established  when $V_\mu$ was absent
connecting self-dual models with topologically massive models and
summarized in Figure 1. By incorporating the $V_\mu$ field, we have
extended these connections so as to include models which were
discussed in the context of Supergravity in ref.\cite{KLRV}. The new
connections are now displayed in Figure 2.

~

Up to this point our analysis was presented at the level of the
partition functions of the different models. We shall end this work
by briefly explaining how we can incorporate external sources to
promote these connections to the level of generating functionals of
Green's functions. In particular, we shall be able to establish that
the Green's functions for $A_\mu $ in the self-dual model with
partition function (\ref{fig}) are those for $\varepsilon_{\mu \nu
\alpha} F_{\nu \alpha}[b]$ in the topologically massive model with
partition function (\ref{entra}), showing thus the following
connection :
\beq
A_\mu  \to  \varepsilon_{\mu \nu \alpha} F_{\nu \alpha}[b]  \; .
\label{nun1}
\eeq

To this end, let us consider the following path-integral :
\beq
Z[s]=
\int \D b_\mu \; \delta[b_\mu - {\tilde b}_\mu(s)] \,
\exp(\frac{1}{2\pi} S_{CS}[b] )
\label{n1}
\eeq
where  ${\tilde b}_\mu(s)$ is defined through
\beq
 \varepsilon_{\mu \nu \alpha} F_{\nu\alpha}[{\tilde b}(s)] =
s_\mu .
\label{n2}
\eeq
After exponentiating the delta functionals by introducing as in the
sourceless case a Lagrange multiplier $A_\mu$, introducing ghosts for
the Faddeev-Popov determinant and auxiliary fields, and using BRST
transformations to write  $A_\mu s_\mu = (\delta \bar c_\mu )s_\mu$,
 one ends with the analogous of eqs.(\ref{20}, \ref{Z})
\beq
Z[s]  =
\int \D b_\mu \D A_\mu \D \bar c_\mu \D c_\mu
 \exp (-\tilde S_{eff}[b,A, \bar c, c\,;s])
\label{n13}
\eeq
with
\begin{eqnarray}
 \tilde S_{eff}[b,A,\bar c, c\,;s] & = &
 -\frac{1}{2\pi} S_{CS}[b] - \frac{i}{\pi} tr \int \dv A_\mu
 {^*F}_\mu[b]  +  \frac{i g^2}{\pi} tr  \int \dv A_\mu A_\mu +
\nonumber \\ & & \frac{2i}{\pi} tr \int \dv \bar c_\mu ({^*D}_{\mu
\alpha}[b] - {g^2} \delta_{\mu \alpha}) c_\alpha +\frac{i}{\pi}tr
\int d^3x \; s_\mu A_\mu
\label{n20}
\end{eqnarray}
As in the sourceless case, if we ``integrate (\ref{n13}) over
$b_\mu$'', i.e. adding (before integration over the auxiliary fields)
the BRST exact form $\delta {\cal H}$ as in (\ref{sui}) and
performing the shift $b \to b - A$, we end with the generating
functional for the self-dual model :
\beq
Z[s]=  \int \D A_\mu \exp(-S_{SD}[A] +\frac{i}{\pi} tr \int d^3x
\; s_\mu A_\mu ) \label{nfig}
\eeq
If instead we integrate (\ref{n13}) over
$A_\mu$  we end with a generating functional for the
topologically massive model :
\begin{eqnarray}
&&Z[s]=\int \D b \D \bar c \D c \; \exp\left( \right.\frac{1}{2\pi}
S_{CS}[b] + \frac{i}{2\pi g^2} tr \int \dv F_{\mu \nu}^2 [b]- \ \ \ \ \ 
\ \ \ \ \ \ \ \ \ \ \ \ \ \ \ \ 
 \nonumber \\ 
 &&\frac{2i}{\pi} tr \int \dv \bar c_\mu
 ({^*D}_{\mu \alpha}[b] - {g^2} \delta_{\mu \alpha} ) c_\alpha
 - \frac{i}{2\pi g^2} tr \int d^3x \; s_\mu ({^*F}_\mu [b] +
 \frac{s_\mu }{2}) \left. \right) 
 \label{nentra}
\end{eqnarray}
The equivalence between (\ref{nfig}) and (\ref{nentra}) implies by
differentiation the connection (\ref{nun1}) up to a coefficient. The
same procedure can be followed in order to add sources to the model
defined by the partition function (\ref{agre}) which includes the
supplementary field $V_\mu$.  We shall not repeat the steps here.

In conclusion: different descriptions of massive self-dual vector
field theory in three dimensions can be linked together with the
aid of a general BRST symmetry. Moreover, all of these models
are descendent of a primal topological field theory, related to the
classical limit of the CS theory. The introduction of external sources
allows us to find the precise correspondence between the basic fields
in the different theories.

~

\underline{Acknowledgments}: F.A.S. and C.N.  are partially supported
by Fundacion Antorchas, Argentina and a Commission of the European
Communities contract No:C11*-CT93-0315. E.M. is partially supported
by CUNY Collaborative Incentive Grant 991999.

\newpage
{\footnotesize

\begin{picture}(140,220)(-30,20)
\put(100,200){\makebox(100,30)
{$Z= \int \D b_\mu \D g \exp(\frac{1}{2\pi} S_{CS}[b] ) \delta[b - 0^g]$}
}
\put(100,170){\makebox(100,30)
{\vector(0,-20){30}}
}
\put(95,170){\makebox(100,30)
{\vector(0,20){30}}
}
\put(160,180){\makebox(10,10)
{$\delta {\cal G}$}
}
\put(220,160){\makebox(10,10)[b]
{$\int \! \D A$}
}
\put(120,140){\makebox(50,30)
{$Z= \int \! \D \Phi \exp(-\tilde S_{eff} )$}
}
\put(210,142){\makebox(30,20)[t]
{\vector(1,0){20}}
}

\put(290,138){\makebox(50,30)
{$Z= \int \! \D b \exp(- S_{top} )$}
}
\put(100,110){\makebox(100,30)
{\vector(0,-20){30}}
}
\put(95,110){\makebox(100,30)
{\vector(0,20){30}}
}
\put(160,120){\makebox(10,10)
{$\delta {\cal H}$}
}
\put(60,105){\makebox(10,10)[b]
{"$\int \! \D b$"}
}
\put(120,80){\makebox(50,30)
{$Z= \int \! \D \Phi \exp(-{\hat S}_{eff} )$}
}
\put(50,80){\makebox(30,30)
{\vector(-1,0){20}}
}
\put(-40,80){\makebox(50,30)
{$Z= \int \! \D A \exp(- S_{SD} )$}
}
\put(-40,40){\makebox(50,30)
{\it  Self-Dual Models }
}
\put(290,40){\makebox(50,30)
{\it  Topologically Massive Models }
}
\put(120,5){\makebox(50,30)
{ \rm Figure 1 }
}
\put(100,-10){\makebox(100,30)
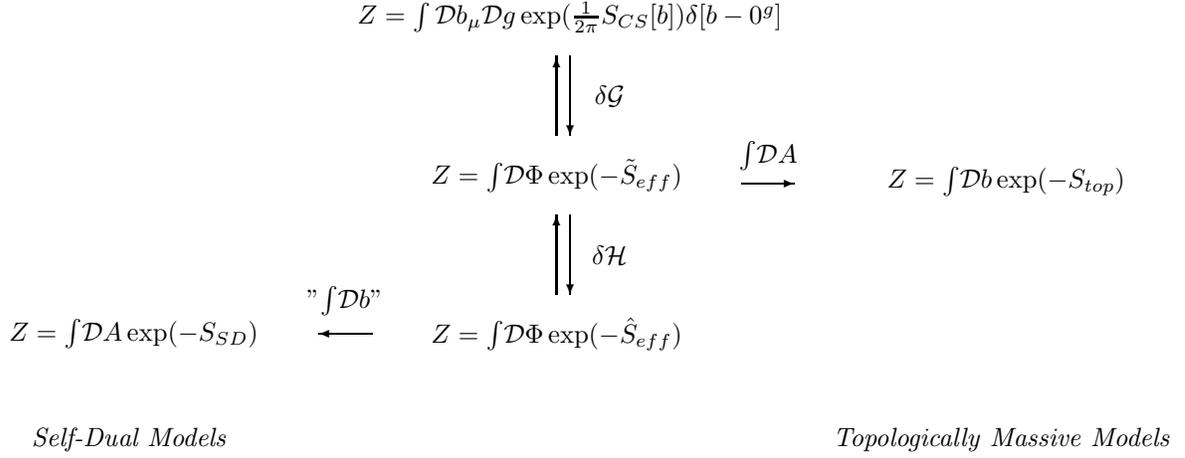
{ \it Connections between different models.}
}

\end{picture}
\vspace{2 cm}

\begin{picture}(140,220)(-30,20)
\put(100,200){\makebox(100,30)
{$Z= \int \! \D \Phi \exp(-{\tilde S}_{eff} - S[V] )$}
}
\put(100,170){\makebox(100,30)
{\vector(0,-20){30}}
}
\put(95,170){\makebox(100,30)
{\vector(0,20){30}}
}
\put(160,180){\makebox(10,10)
{$\delta {\cal K}(\lambda)$}
}
\put(220,160){\makebox(10,10)[b]
{$\lambda = 0, \int \! \D A$}
}
\put(120,140){\makebox(50,30)
{$Z= \int \D \Phi \exp(-S_{eff})$}
}
\put(210,142){\makebox(30,20)[t]
{\vector(1,0){20}}
}

\put(290,138){\makebox(50,30)
{$Z= \int \! \D b \D V \exp(- S[b,V] )$}
}
\put(68,163){\makebox(10,10)[b]
{$ \lambda=\frac{2}{3}$, "$\int \! \D b $"}
}
\put(58,140){\makebox(30,30)
{\vector(-1,0){20}}
}
\put(-40,140){\makebox(50,30)
{$Z= \int \! \D A \D V \exp(- {S}[A,V])$}
}
\put(-40,40){\makebox(50,30)
{\it  Self-Dual Models }
}
\put(290,40){\makebox(50,30)
{\it  Topologically Massive Models }
}
\put(120,10){\makebox(50,30)
{ \rm Figure 2 }
}
\put(100,-10){\makebox(100,30)
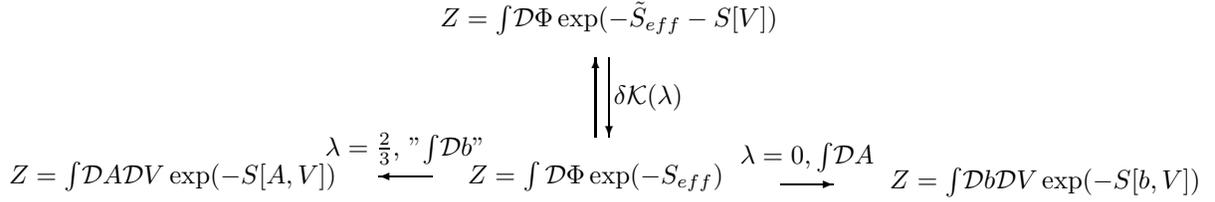
{ \it New connections when the vector field $V_\mu$ is considered.
}
}

\end{picture}
\vspace{2 cm}
}

\end{document}